\documentclass[twocolumn,aps,showpacs,superscriptaddress]{revtex4}
\usepackage{amsmath,amssymb,eucal,graphicx}
\begin{document}
\title{Polymerization with Freezing}
\author{E.~Ben-Naim}
\affiliation{Theoretical Division and Center for Nonlinear
Studies, Los Alamos National Laboratory, Los Alamos, New Mexico
87545, USA}
\author{P.~L.~Krapivsky}
\affiliation{Center for Polymer Studies and Department of Physics,
Boston University, Boston, Massachusetts 02215, USA}

\begin{abstract}
  
  Irreversible aggregation processes involving reactive and frozen
  clusters are investigated using the rate equation approach. In
  aggregation events, two clusters join irreversibly to form a larger
  cluster, and additionally, reactive clusters may spontaneously
  freeze. Frozen clusters do not participate in merger events.
  Generally, freezing controls the nature of the aggregation process,
  as demonstrated by the final distribution of frozen clusters.  The
  cluster mass distribution has a power-law tail, $F_k\sim
  k^{-\gamma}$, when the freezing process is sufficiently
  slow. Different exponents, $\gamma=1$, $3$ are found for the
  constant and the product aggregation rates, respectively. For the
  latter case, the standard polymerization model, either no gels, or a
  single gel, or even multiple gels may be produced.

\end{abstract}


\maketitle

\section{Introduction} 

Aggregation is a fundamental irreversible process in which physical
objects merge irreversibly to form larger objects. Aggregation has
numerous applications ranging from astronomy where planetary systems
form via collisions of planetesimals, to atmospheric science
\cite{drake,sein}, to chemical physics where polymeric chains
chemically bond and form polymeric networks or gels
\cite{flory,stock,floryB}, to computer science
\cite{Aldous,jlr,bk-rev}.

The standard framework for modeling aggregation is as
follows. Initially, the system consists of a large number of identical
molecular units (``monomers''). A cluster (``polymer'') is composed of
an integer number of monomers, termed the cluster mass. In each
aggregation event, a pair of clusters merges, thereby forming a larger
cluster whose mass equals the sum of the two original masses.

When the number of aggregation events is unlimited, the system
condenses into a single cluster. However, in most practical
applications, other processes intervene well before that, and as a
result, the final state has multiple clusters, rather than a single
condensate. For example, fragmentation of large clusters into smaller
clusters is one mechanism that may counterbalance aggregation and
prevent condensation.

In this study, we focus on another control mechanism: freezing. We use
the generic term freezing to describe situations where there are two
types of clusters: reactive clusters that participate in aggregation
events and passive clusters that do not participate in aggregation
events.

In particular, we consider the case where reactive clusters have a
finite lifetime. In our model, reactive clusters spontaneously turn
into frozen clusters.  Spontaneous freezing can occur via various
mechanisms. For instance, the environment may contain ``traps'' that
absorb the diffusing polymers. In this situation, the reactive
clusters are the free (mobile) polymers and the frozen clusters are
the polymers adsorbed to the trap surface.  Another example is a
system of linear polymers with reactive end monomers. In a merging
event, two different chains chemically bond via the end monomers,
while in a freezing event, the two end monomers of the same chain bond
to form a ring. Rings can no longer participate in aggregation
events. Thus, in this case the linear polymers are reactive and the
ring polymers are frozen.

In aggregation with freezing, it is natural to consider the initial
condition where there are reactive clusters only. Of course, the
system ends with frozen clusters only. Of special interest is the
final mass distribution of frozen clusters.  In this study, we address
the two classical aggregation rates.

First, we study the simplest aggregation process where the aggregation rate
is independent of the cluster mass. We find that the mass distribution of
reactive clusters decays exponentially with the cluster mass. In general, the
mass density of frozen clusters also decays exponentially with the cluster
mass. However, when the freezing rate is very small, there is a power-law
behavior, $F_k\sim k^{-1}$, over a substantial range of masses.

Second, we consider the case where the aggregation rate is
proportional to the product of the masses, a process that is widely
used to model polymerization and gelation. We find that when the
freezing rate exceeds a certain threshold, no gels form, while when
the freezing rate is below this threshold, at least one gel
forms. Interestingly, the number of gels produced fluctuates from
realization to realization. For supercritical freezing rates, the mass
distribution of frozen clusters decays exponentially, while below this
threshold, it decays algebraically, $F_k\sim k^{-3}$ \cite{bk}.
 
\section{The master equations} 

We analyze the stochastic process of aggregation with freezing using
the rate equation approach. Let us first consider the evolution of
reactive clusters. In aggregation processes, two reactive clusters of
masses $i$ and $j$ merge to form a larger reactive cluster of mass
$i+j$. The aggregation rate $K(i,j)$ is a function of the two cluster
masses. The freezing process is random: reactive clusters may
spontaneously freeze with a constant rate.  This freezing rate $f_k$
may be mass-dependent. Therefore, the mass distribution $R_k(t)$ of
reactive clusters of mass $k$ at time $t$ evolves according to the
generalized Smoluchowki equation
\begin{eqnarray}
\label{Rk} 
\frac{dR_k}{dt}\!=\!\frac{1}{2}\sum_{i+j=k}K(i,j)R_iR_j
\!-\!R_k\sum_{i=1}^\infty K(i,k)R_i\! -\! f_k R_k.
\end{eqnarray}
The first two terms account for gain and loss of clusters of mass $k$,
and the last term accounts for loss due to freezing. This master
equation assumes perfect mixing as the probability of finding two
clusters at the same position is a product of the probabilities of
finding each of the clusters independently at the same position. We
restrict our attention to the natural case of monodisperse initial
condition, $R_k(0)=\delta_{k,0}$.

The mass distribution of frozen clusters $F_k(t)$ is coupled to the
mass distribution of reactive clusters according to the rate equation
\begin{equation}
\label{Fk} 
\frac{dF_k}{dt}=f_k R_k.
\end{equation}
It is simple to check that the total mass density $\sum_k
k(R_k+F_k)=1$ is conserved by the evolution equations
(\ref{Rk})--(\ref{Fk}). Initially, there are no frozen clusters so
$F_k(0)=0$. Eventually, all clusters become frozen, so the final mass
distribution $F_k(\infty)$ of frozen clusters is of particular
interest.

The master equations (\ref{Rk})--(\ref{Fk}) are sets of infinitely
many coupled nonlinear differential equations, and they are generally
unsolvable. Even in the absence of freezing, these equations are
solvable only for special aggregation rates
\cite{drake,Aldous,fran}. The three classical solvable cases are the
constant rate $K(i,j)={\rm const.}$, the sum rate $K(i,j)=i+j$, and
the product rate $K(i,j)=ij$. These cases represent natural
aggregation processes. Mass independent aggregation rates correspond
to an aggregation process where two clusters are chosen randomly to
merge. Aggregation rates proportional to the product of the two
cluster masses correspond to polymerization processes where two
monomers are picked randomly to form a chemical bond; consequently,
their respective clusters are merged. The sum rate is a hybrid between
the two as it is an aggregation process where a randomly chosen
monomer bonds with a randomly chosen polymer. In this study, we focus
on the two most widely used cases of constant and product aggregation
rates.

\section{Constant Aggregation Rate} 

First, we discuss how the constant aggregation rate relates to
polymerization in the presence of traps. To treat this problem
formally, one should write down the master equations with
inhomogeneous densities and add diffusion terms. Then, one should
study these equations in the trap-free region subject to the absorbing
boundary conditions imposed by the traps. This approach is not
practical and the reaction-rate approach provides a powerful
alternative \cite{smol,chandra,ovchin,gleb,redner}.  The reaction-rate
approach is roughly speaking an effective-medium theory that ignores
the complicated influence of each trap on the diffusion of particles
and instead, represents this influence by averages.  The reaction-rate
approach was used by Smoluchowski to compute the aggregation rate
$K(i,j)$ for Brownian particles. Assuming that merging happens
immediately upon collision, and that particles are spherical and have
radii $R_i$ and $R_j$ and diffusion coefficients $D_i$ and $D_j$,
Smoluchowski obtained
\begin{equation}
\label{K} 
K(i,j)=4\pi (D_i+D_j)(R_i+R_j).
\end{equation}
Stokes' law shows that the diffusion coefficient of a Brownian
particle is inversely proportional to its radius, $D_k\sim 1/R_k\sim
k^{-1/3}$, and therefore the Brownian kernel becomes
\begin{equation}
\label{KB} 
K(i,j)\propto \left(i^{-1/3}+j^{-1/3}\right)\left(i^{1/3}+j^{1/3}\right).
\end{equation}
Here, we ignored the overall multiplicative factor as it is irrelevant
for the current discussion.

The master equations with this complicated Brownian kernel have not
been solved even in the case of pure aggregation. To simplify the
analysis, Smoluchowski suggested to replace the Brownian kernel
(\ref{KB}) by the constant kernel. These two kernels have one common
feature---they both are invariant under the dilatation
$K(ai,aj)=K(i,j)$. Therefore, one expects that both kernels lead to
similar behaviors, and to a certain extent, i.e., as far as overall
scaling properties are concerned, this approximation is sensible
\cite{fran}.
 
A straightforward extension of Eq.~(\ref{K}) gives the freezing rate 
\begin{equation}
\label{fk} 
f_k=4\pi\,n\, (D_k+D)(R_k+a)
\end{equation}
where $n$ is the density of traps that are assumed to be spheres of
radius $a$ and diffusion coefficient $D$. The clusters are usually
polymers whose molecular weight is small compared to the size of the
traps; hence $R_k\ll a$ and $D_k\gg D$. Therefore $f_k=4\pi a nD_k$,
yielding the mass dependence
\begin{equation}
\label{fk1} 
f_k\propto k^{-1/3}.
\end{equation}

Thus, a constant aggregation rate together with spontaneous freezing
approximate aggregation of Brownian particles in the presence of
traps. We stress that the use of a constant aggregation rate instead
of (\ref{KB}) is an approximation.

\subsection{Constant freezing rates} 

Since the constant aggregation rate merely sets the overall time
scale, we may conveniently set its value $K(i,j)=2$ without loss of
generality. Let us first consider constant freezing rates,
$f_k=\alpha$. The master equation (\ref{Rk}) becomes
\begin{equation}
\label{Rk-const} 
\frac{dR_k}{dt}=\sum_{i+j=k}R_iR_j-(2R+\alpha)R_k. 
\end{equation}
Here, we used the total density of reactive clusters, $R=\sum_k
R_k$. In general, for mass-independent freezing rates, it is possible
to eliminate the freezing term from the master equation by
transforming the mass distribution $R_k=C_k\, e^{-\alpha t}$ and
introducing the time variable 
\begin{equation}
\label{tau} 
\tau=\frac{1-e^{-\alpha t}}{\alpha}\,. 
\end{equation}
The time variable $\tau$ grows from $0$ to $\alpha^{-1}$ as the
physical time increases from $0$ to $\infty$.  With these
transformations, the governing equations for the densities $C_k$
reduce to the pure aggregation case
\hbox{$dC_k/d\tau=\sum_{i+j=k}C_iC_j-2C\,C_k$} with the total density
$C=\sum_k C_k$. We briefly recall how to solve these equations. The
total density obeys $dC/d\tau=-C^2$ and subject to the initial
condition $C(0)=1$, the total density is $C(\tau)=(1+\tau)^{-1}$. Let
us now introduce the exponential ansatz $C(\tau)=A\,a^{k-1}$ with
$A(0)=1$ and $a(0)=1$ to satisfy the initial conditions. Substituting
this ansatz into the master equation and equating mass-independent and
mass-dependent terms separately, yields $dA/d\tau=-2(1+\tau)^{-1}A$
and therefore, $A=(1+\tau)^{-2}$, and $da/d\tau=A$ leading to
$a=\tau/(1+\tau)$. The well-known solution for the pure aggregation
case is therefore
\begin{equation}
\label{Ck-sol} 
C_k(\tau)=\frac{\tau^{k-1}}{(1+\tau)^{k+1}}\,.
\end{equation}

Thus, the mass distribution of reactive clusters reads
\begin{equation}
\label{Rk-sol} 
R_k(\tau)=(1-\alpha\tau) \frac{\tau^{k-1}}{(1+\tau)^{k+1}}\,. 
\end{equation}
The exponential mass dependence is as in the pure aggregation
case. Also, the total density of reactive clusters is $R=(1-\alpha
\tau)/(1+\tau)$ and as expected, the reactive clusters do eventually
deplete $R(t=\infty)=R(\tau=1/\alpha)=0$.

The mass distribution of frozen clusters is found by integrating the
equation $dF_k/d\tau=\alpha\,C_k$ with respect to time.  Substituting
(\ref{Ck-sol}), and using \hbox{$d\tau/dt=(1-\alpha\tau)=e^{-\alpha
t}$}, the integration is immediate and
\begin{equation}
\label{Fk-sol} 
F_k(\tau)=\frac{\alpha}{k}\left(\frac{\tau}{1+\tau}\right)^k.
\end{equation}
We see that in addition to the dominant exponential behavior, there is
an additional algebraic prefactor. The total density of frozen
clusters $F=\sum_k F_k$ is found by summation, $F(\tau)=\alpha\ln
(1+\tau)$ and in particular, the final density of frozen clusters is
$F(\infty)\equiv F(t=\infty)=\alpha\ln (1+1/\alpha)$. Also, 
the final mass distribution of frozen clusters is 
\begin{equation}
\label{Fk-final} 
F_k(\infty)=\frac{\alpha}{k}\left(\frac{1}{1+\alpha}\right)^k.
\end{equation}
In general, the mass distribution decays exponentially, but there is
a $k^{-1}$ algebraic correction. 

\subsection{Slow freezing} 

The most interesting behavior occurs in the slow freezing limit: as 
$\alpha\to 0$, the final mass distribution becomes algebraic
\begin{equation}
F_k(\infty)\simeq \alpha\,k^{-1}. 
\end{equation}
This power law holds over a substantial mass range, \hbox{$k\ll
\alpha^{-1}$}. Beyond this scale, the tail is exponential,
\hbox{$F_k(\infty)\simeq \alpha\,k^{-1}\,e^{-\alpha k}$}.

The results in the slow freezing limit can be alternatively obtained
using perturbation theory. Indeed, the modified time variable
coincides with the original time variable, $\tau\to t$ as $\alpha\to
0$, and the pure aggregation results are recovered. In other words,
the freezing loss term $-f_kR_k$ can be neglected in the master
equation (\ref{Rk}). Using this perturbation approach we address two
related problems: general freezing rates and aggregation in
low-dimensional systems.

Let us consider general mass dependent freezing rates $f_k$. Dropping
the loss rate from the master equation, the reactive cluster density
is as in the pure aggregation case $R_k=t^{k-1}(1+t)^{-k-1}$, given by
Eq.~(\ref{Ck-sol}). The mass distribution of frozen clusters is
obtained by integrating Eq.~(\ref{Fk})
\begin{equation}
\label{Fkt-sol} 
F_k(t)=\frac{f_k}{k}\left(\frac{t}{1+t}\right)^k.
\end{equation}
We see that the algebraic prefactor $k^{-1}$ is generic. Therefore,
the final mass distribution is
\begin{equation}
\label{Fk-inf} 
F_k(\infty)=k^{-1}\,f_k.
\end{equation}
This behavior applies for masses below some threshold mass $k_*$,
while the mass distribution sharply vanishes above the threshold. The
threshold mass is estimated from mass conservation:
\begin{equation} 
\label{threshold} 
1=\sum_{k\geq 1} kF_k(\infty)\sim \sum_{k=1}^{k_*} f_k.
\end{equation}
As argued above, for Brownian coagulation in the presence of traps,
$f_k=\beta k^{-1/3}$. For slow freezing, $\beta\ll 1$, we conclude
that the final mass distribution is algebraic, $F_k(\infty)\simeq \beta
k^{-4/3}$, below the threshold mass $k_*\sim \beta^{-3/2}$.

The rate equation approach neglects spatial correlations as the
probability of finding two clusters at the same position is
represented by the product of the probabilities of finding each cluster
separately at the same position. This mean-field approximation is
valid only when the spatial dimension exceeds the critical dimension
$d_c$ \cite{gleb,redner}. It is therefore interesting to study the
behavior below the critical dimension. 

We address here the Point Cluster Model (PCM) where the radii and the
diffusion coefficients are both mass-independent. In this case
$d_c=2$.  The lattice PCM is defined as follows: clusters occupy
single lattice sites and hop to adjacent sites with rate $D$; if a
reactive cluster hops onto a site occupied by another reactive
cluster, both clusters merge. We assume that frozen clusters do not
affect reactive clusters.  The PCM without freezing can be solved
exactly in one dimension. When all lattice sites are initially
occupied by monomers, the density of reactive clusters of mass $k$ is
\cite{spouge,af}
\begin{equation}
\label{spouge-sol} 
R_k(t)=e^{-4Dt}\left[I_{k-1}(4Dt)-I_{k+1}(4Dt)\right]
\end{equation}
where $I_n$ is the modified Bessel function of order $n$.  Here, we
implicitly considered the slow freezing limit. The density of frozen
clusters is found from $dF_k/dt=f_k R_k$, that is of course always
valid. Using the identity \hbox{$\int_0^\infty
dx\,e^{-x}\left[I_{k-1}(x)-I_{k+1}(x)\right]=2$}, the final
distribution of frozen clusters is
\begin{equation}
\label{Fk-inf-1D} 
F_k(\infty)=(2D)^{-1}\,f_k.
\end{equation}
Remarkably, the very same answer (\ref{Fk-inf-1D}) is also found for
the continuous version of the PCM. The mass distribution
(\ref{Fk-inf-1D}) holds up to a certain threshold mass. For example,
for the constant freezing rate $f_k=\alpha\ll 1$ the threshold mass is
$k_*\sim \sqrt{D/\alpha}$.

\section{Product aggregation rate} 

The product aggregation rate $K(i,j)=ij$ is equivalent to the
Flory-Stockmayer gelation model \cite{flory,stock,aal}. In this
model, any two monomers may form a chemical bond and when this happens
the two respective polymers become one. Thus, the aggregation rate
equals the product of the cluster masses. In this polymerization
process, a polymer network (``gel'') emerges in a finite time, and it
is giant in the sense that it contains a finite fraction of the
monomers in the system. Eventually it grows to engulfs the entire
system. This gelation model is also the simplest mean-field model of
percolation \cite{zhe,sh}.

As in the previous section, we analyze in detail mass-independent
freezing rates, $f_k=\alpha$, for which the master equation (\ref{Rk})
becomes
\begin{equation}
\label{Rkt} 
\frac{dR_k}{dt}=\frac{1}{2}\sum_{i+j=k}ij R_i R_j - m k R_k - \alpha R_k
\end{equation}  
where $m$ is the total mass density of reactive clusters. If all
clusters are finite in size then $m=M_1=\sum_{k\geq 1}kR_k$ where
$M_n=\sum_k k^n R_k$ is the general $n$th moment of the
distribution. Again, we consider the monodisperse initial conditions 
$R_k(t)=\delta_{k,1}$ and $F_k(0)=0$.

Low order moments of the mass distribution obey closed equations and
thus, provide a useful probe of the aggregation dynamics. The total
mass density of reactive clusters satisfies $dm/dt=-\alpha m$
reflecting the loss due to freezing, and therefore
\begin{equation}
\label{M1-sol}
m(t)=m(0)e^{-\alpha t}\,.
\end{equation}
The total mass density decays exponentially with the physical time, or
equivalently, linearly with the modified time, $m(\tau)=1-\alpha
\tau$. Furthermore, the second moment includes in addition to the
linear loss term, a nonlinear term that accounts for changes due to
aggregation, $dM_2/dt=M_2^2-\alpha M_2$.  Solving this equation with
arbitrary initial condition yields
\begin{equation}
\label{M2-sol}
M_2(t)=\alpha\,\left[\left(\frac{\alpha}{M_2(0)}-1\right)\,
e^{\alpha t}+1\right]^{-1}.
\end{equation}
Divergence of the second moment signals the emergence of a gel in a
finite time, i.e., the occurrence of the gelation phase
transition. Fixing the freezing rate, the initial conditions govern
whether gelation does or does not occur: gelation occurs when the
initial mass is sufficiently large, $M_2(0)>\alpha$, but otherwise
there is no gelation. Conversely, fixing the initial conditions,
gelation occurs only for slow enough freezing. For the monodisperse
initial conditions, the critical freezing rate is $\alpha_c=1$. When
gelation does occur, the gelation time is
\begin{equation}
\label{tg}
t_g=-\frac{1}{\alpha}\ln \left(1-\frac{\alpha}{M_2(0)}\right).
\end{equation}
The gelation point separates two phases. Prior to the gelation time,
the system contains only finite clusters that undergo cluster-cluster
aggregation. Past the gelation point, the gel grows via cluster-gel
aggregation. We term these two the coagulation phase and the gelation
phase, respectively. The above expressions for the first two moments
are valid for the coagulation phase only.

The mass distribution of reactive clusters is found again by
transforming the mass distribution $R_k=e^{-\alpha t}C_k$ and the time
variable (\ref{tau}). With these transformations, the problem reduces
to the no-freezing case, \hbox{$dC_k/dt=\frac{1}{2}\sum_{i+j=k} ij C_i
C_j-kMC_k$} with $m=Me^{-\alpha t}$. Using the variable
$u(\tau)=\int_0^\tau M(\tau') d\tau'$ and the transformation $C_k=G_k
\tau^{k-1}e^{-ku}$, the master equation reduces to a recursion
equation for the {\it time-independent} coefficients $G_k$:
$(k-1)G_k=\frac{1}{2}\sum_{i+j=k}ij G_iG_j$. This equation is solved
using the generating function technique. The so-called
``tree-function'' $G(z)=\sum_k kG_k e^{kz}$ satisfies $dG/dz=G/(1-G)$
and the solution of this differential equation obeys
$Ge^{-G}=e^z$. The coefficients $G_k=k^{k-2}/k!$ are found using the
Lagrange inversion formula \cite{wilf}. Hence, the mass distribution
of reactive clusters is
\begin{equation}
\label{formal}
R_k(\tau)=\frac{k^{k-2}}{k!}\,\,(1-\alpha\tau)\,\tau^{k-1}\,e^{-k\,u}\,.
\end{equation}
The corresponding generating function
$\mathcal{R}(z)=\sum_{k=1}^\infty k R_k e^{kz}$ can be expressed in
terms of the tree-function 
\begin{equation}
\label{rzt}
\mathcal{R}(z)=\tau^{-1}(1-\alpha\tau)\,G(z+\ln \tau-u).
\end{equation}
Explicitly, the tree function is $G(z)=\sum_{k\geq
1}\frac{k^{k-1}}{k!}\, e^{kz}$.

The mass distribution (\ref{formal}) is only a formal solution because
the total mass density $m$ and hence the variable $u$ are yet to be
determined. Prior to gelation, the solution can be obtained in an
explicit form because the various variables are known.  From the first
moment (\ref{M1-sol}) then $M=1$ and therefore
\begin{equation}
\label{u}
u=\tau
\end{equation}
for $t<t_g$. In this case, the mass distribution decays exponentially
at large masses and the typical cluster mass is finite. When
$\alpha>1$, there is no gelation transition, and this behavior
characterizes the mass distribution at all times. Otherwise, when
gelation does occur, the gelation time (\ref{tg}) is simply
$\tau_g=1$.  The gelation point is marked by an algebraic divergence
of the mass distribution, $R_k(t_g)\sim (1-\alpha)k^{-5/2}$, for large
$k$.

Using the explicit expression for $R_k$ prior to gelation, we can
calculate the mass distribution of the frozen clusters produced up to
that point. Substituting (\ref{u}) into the formal solution
(\ref{formal}) and integrating $dF_k/dt=\alpha R_k$ over time using
$d\tau/dt=e^{-\alpha t}=(1-\alpha \tau)$ yields the distribution $F_k(t_g)$
of frozen clusters produced prior to gelation ($t_g\equiv\infty$ for
$\alpha>1$)
\begin{equation}
\label{Pk-final}
F_k(t_g)=
\begin{cases}
\frac{\alpha}{k^2\cdot k!}\,\,\gamma(k,k)&\alpha\leq 1,\\
\frac{\alpha}{k^2\cdot k!}\,\,\gamma(k,k/\alpha)&\alpha\geq 1,
\end{cases}
\end{equation}
where $\gamma(n,x)=\int_0^x dy\,y^{n-1}\,e^{-y}$ is the incomplete
gamma function. When $\alpha\geq 1$, this quantity equals the final
distribution of frozen clusters, $F_k(\infty)=F_k(t_g)$. At large
masses, the behavior is as follows
\begin{equation}
\label{fk-tail}
F_k(\infty)\simeq
\begin{cases}
\frac{1}{2}\cdot k^{-3}                          &\alpha=1,\\
A(\alpha)\,k^{-7/2}\exp\left[-B(\alpha) k\right] &\alpha>1,
\end{cases}
\end{equation}
where $A=(2\pi)^{-1/2}\alpha^2/(\alpha-1)$ and
$B=\alpha^{-1}+\ln\alpha -1$. These asymptotic results were obtained
using the steepest descent method. Quantitatively, the mass
distribution differs from that obtained for constant aggregation
rates. However, qualitatively, there is a similarity: there is
exponential decay above a critical freezing rate and algebraic decay
at and below this critical freezing rate. For the constant aggregation
rate, the critical freezing rate vanishes, but for the product
aggregation rate the critical rate is finite.

\begin{figure}[t]
 \includegraphics*[width=0.4\textwidth]{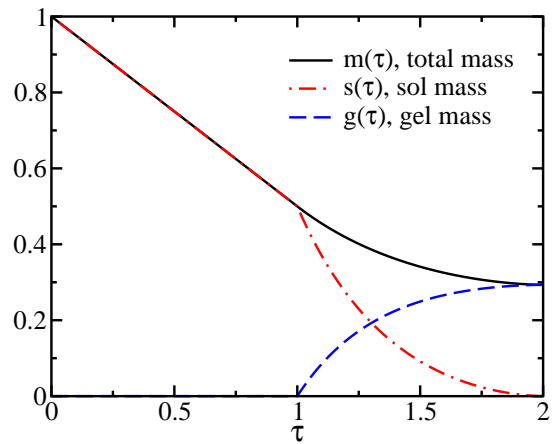}
 \caption{The total mass, the sol mass, and the gel mass versus
the modified time $\tau$ for $\alpha=1/2$.}
\label{sol} 
\end{figure}

At the gelation transition a giant cluster that contains a fraction of
the mass in the system emerges. Past the gelation point, two
aggregation processes occur in parallel: in addition to
cluster-cluster aggregation, the giant cluster grows by swallowing
finite clusters. Now, reactive clusters consist of finite clusters
(the ``sol'') with mass $s=M_1$, and the gel with mass $g$.  The total
mass density is $m=s+g$. These three masses are coupled via the
evolution equations
\begin{subequations}
\label{mst-eq}
\begin{align}
\frac{dm}{dt}&= - \alpha\,s,\\
\frac{ds}{dt}&=-\alpha\,s-\frac{s(m-s)}{1-s\tau e^{\alpha t}}.
\end{align}
\end{subequations}
The initial conditions are $m(t_g)=s(t_g)=1-\alpha$.  The first
equation reflects that as long as the gel remains reactive, mass may
be converted from the reactive state to the frozen state via freezing
of finite clusters.  The second equation follows from $ds/dt=-\alpha
M_1-g\,M_2$, obtained by summing (\ref{Rkt}). The second moment is
written explicitly, $M_2= \mathcal{R}_z(z=0)=s/(1-s\tau e^{\alpha
t})$, using the aforementioned identity \hbox{$G'(z)=G/(1-G)$}.  The
initial conditions are $m(t_g)=s(t_g)=1-\alpha$. Once these equations
are solved, the solution (\ref{formal}) becomes explicit. We analyze
this equation using perturbation theory in the limits $\alpha\uparrow
1$ and $\alpha\downarrow 0$ as detailed in Appendix A.  For general
freezing rates, we solve these equations numerically (Fig.~\ref{sol}).

In addition to the freezing of the finite clusters, the gel itself may
freeze. One way to characterize the gel by its maximal possible size
$g_{\rm max}=\lim_{t\to\infty}g(t)$. The limiting behaviors are as
follows (see Appendix A)
\begin{equation}
\label{g-pert} 
g_{\rm max}\to 
\begin{cases}
1-\frac{\pi^2}{6}\alpha&\alpha \downarrow 0,\\
C(1-\alpha)^2&\alpha \uparrow 1,
\end{cases}
\end{equation}
with $C=1.303892$. The maximal gel size decreases as the freezing rate
increases. Just below the critical freezing rate, the gel is very
small as its size shrinks quadratically with the distance from the
critical point $g\sim (1-\alpha)^2$; perturbation analysis shows that
this behavior is generic and not limited to the maximal gel
size. Therefore, freezing is a mechanism for controlling the gel size:
by using freezing rates just below the critical rate, it is possible
to produce micro-gels.

\begin{figure}[t]
 \includegraphics*[width=0.4\textwidth]{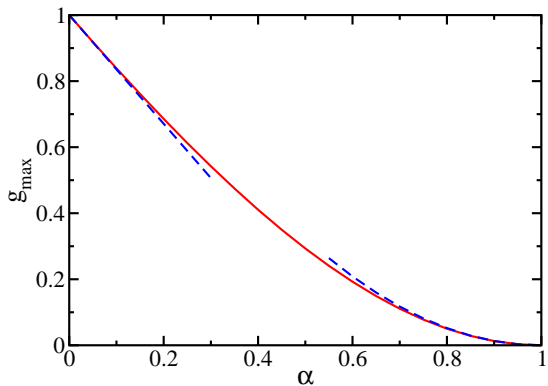}
 \caption{The maximal gel mass $g_{\rm max}$ versus the freezing rate
 $\alpha$ (solid line). Perturbation theory results are shown using 
 dashed lines.}
\label{gmax} 
\end{figure}

\begin{figure}[t]
\includegraphics*[width=0.375\textwidth]{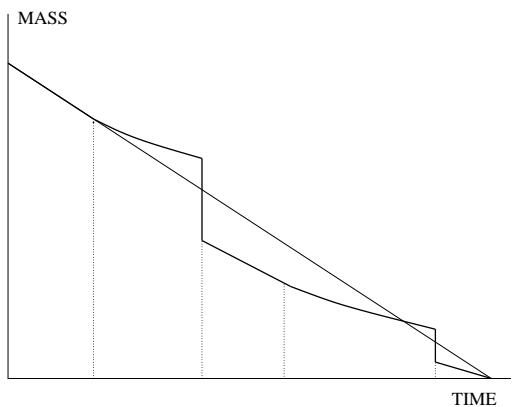}
\caption{The mass density $m$ (MASS) versus time $\tau$ (TIME).  The
 system alternates between the coagulation phase and the gelation
 phase. In the former phase the mass decreases linearly according to
 (\ref{M1-sol}) such that depletion occurs at time $\tau=1/\alpha$. In
 the latter phase, the active mass decreases slower than linear
 according to (\ref{mst-eq}).  The gelation phase ends when the gel
 freezes.}
\label{mass}
\end{figure}

The gel freezes following a random Poisson process: its lifetime $T$
is distributed according to the exponential distribution 
\begin{equation}
\label{pt}
P(T)=\alpha\, e^{-\alpha T}. 
\end{equation}
As long as the gel is active the system evolves in a deterministic
fashion. When the gel freezes, the total reactive mass exhibits a
discontinuous downward jump (Fig.~\ref{mass}). Since the gel freezes
according to a random process, the mass of the frozen gel is also a
random variable. Moreover, this quantity is not self-averaging as it
fluctuates from realization to realization.

When the gel freezes, the system consists of finite-mass clusters only
and therefore, the system re-enters the coagulation phase. The system
may then undergo a second gelation transition that ends when the gel
freezes. Therefore, the evolution is cyclic with the system
alternating between the coagulation and gelation. For the same reason
that the gel mass fluctuates, so is the number of frozen gels produced
a fluctuating quantity.

The number of gels produced is in principle unlimited, i.e., there is
a finite probability $P_n>0$ that $n$ gels are produced. Since the
second moment diverges at the gelation transition according to
(\ref{M2-sol}), it is necessarily larger than the freezing rate at
some finite time interval following the gelation transition. If the
gel freezes during this time interval then a successive gelation is
bound to occur. We also note that the evolution in the coagulation
phase is deterministic and for example, the first two moments follow
Eqs.~(\ref{M1-sol}) and (\ref{M2-sol}). The ``initial'' conditions are
given by the state of the system when the gel freezes.

\begin{figure}[t]
 \includegraphics*[width=0.4\textwidth]{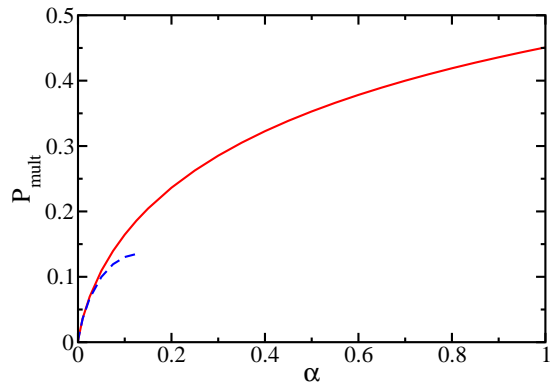}
 \caption{The multiple gel probability $P_{\rm mult}$ versus the freezing rate
 $\alpha$ (solid line). Perturbation theory results are shown using a
 dashed line.}
\label{pm} 
\end{figure}

It is therefore natural to ask: what is the probability that multiple
gels are produced? This is the probability that the gel freezes prior
to time $t_*$ given by $M_2(t_*)=\alpha$. Using the second moment
$M_2=s/(1-s\tau e^{\alpha t})$ this condition simplifies to
\begin{equation}
\label{cond}
s(t_*)=\alpha\, e^{-\alpha t_*}. 
\end{equation}
The probability that multiple gels are produced is obtained by
integrating (\ref{pt}) up to this time, \hbox{$P_{\rm
mult}=\int_0^{t_*-t_g} dT\, P(T)$} with the limiting behaviors (see
Appendix A)
\begin{equation}
\label{pmult} 
P_{\rm mult}\to 
\begin{cases}
\alpha \ln \frac{1}{e\alpha}&\alpha \downarrow 0,\\
0.450851&\alpha \uparrow 1.
\end{cases}
\end{equation}
This probability increases as the freezing rate increases (figure
\ref{pm}).  It is generally substantial, and moreover, it exhibits a
discontinuity at $\alpha=\alpha_c$.

\begin{figure}[t]
 \includegraphics*[width=0.4\textwidth]{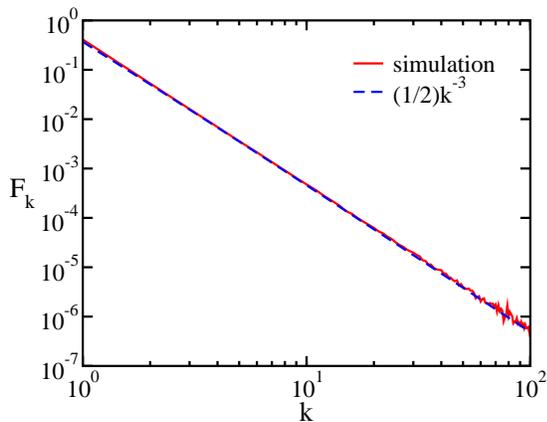}
 \caption{The final mass distribution of the frozen clusters for
   $\alpha=1/2$. The simulation results represent an average over
   $10^2$ independent realizations in a system of mass $N=10^6$.}
\label{frozen}
\end{figure}

We now address the final mass distribution of frozen gels.  Analytic
treatment of the successive gelation phases is difficult due to the
stochastic nature of the freezing process.  Numerically, there are two
ways to treat the problem. One may integrate the rate equations
(\ref{mst-eq}) up to the gel freezing time that is distributed
according to (\ref{pt}) and then repeat this procedure if a successive
gelation occurs. We prefer Monte Carlo simulations where, since the
system is finite, there is no need to distinguish the gel from the
finite clusters.  In the simulations, we keep track of the total
aggregation rate $R_a=N(M_1^2-M_2)/2$ and of the total freezing rate
$R_f=\alpha NM_0$, where $N$ is the number of monomers. Aggregation
occurs with probability $R_a/(R_a+R_f)$, and freezing occurs with the
complementary probability. A cluster is chosen for aggregation with
probability proportional to its mass. Time is augmented by $\Delta
t=1/(R_a+R_f)$ after each aggregation or freezing event.

For the case $\alpha<1$, numerical simulations provide convincing 
evidence that the tail behavior 
\begin{equation}
F_k(\infty)\sim k^{-3}, \qquad \alpha<1
\end{equation}
is universal (Fig.~\ref{frozen}). This indicates that frozen clusters
produced during the coagulation phase dominate at large masses as
every such phase is expected to contribute $k^{-3}$ to the tail
according to Eq.~(\ref{fk-tail}). Intuitively, this is clear because
large clusters are quickly merged into the gel and therefore, frozen
clusters produced in the presence of the gel tend to be small.
Interestingly, freezing leads to a new critical exponent in mean-field
percolation. 

\section{Conclusions}

In summary, we studied aggregation processes with freezing. For
constant freezing rates, the problem can be formally reduced to the no
freezing case. The mass distribution of frozen clusters resembles the
mass distribution of reactive clusters, decaying exponentially at
large masses, when the freezing is sufficiently fast. Novel behavior
emerges when the freezing rate is slower than some critical value. In
this case, the mass distribution of frozen clusters decays
algebraically. For constant aggregation rate, the critical freezing
rate is zero but for the product aggregation rate, it is finite.

For the product aggregation rate, the freezing rate controls the
gelation process. If it is sufficiently high, no gelation occurs, and
if it is just below the threshold, micro-gels are produced. If one gel
is produced, then multiple gels are possible. In this case, the mass
of the gels and their number are both controlled by a random process
and as a result, they fluctuate from realization to realization. The
system exhibits a series of gelation transitions and it alternates
between ordinary coagulation and gelation.  The random freezing
process governs the number of percolation transitions as well as the
mass of the frozen gels.

The behavior when freezing occurs spontaneously is quite different
than that found when freezing is reaction-induced (upon merger a
cluster may freeze with some fixed probability)
\cite{kb,bk-scaling}. For the constant aggregation rate, the mass
distribution is always algebraic and the characteristic exponent is
non-universal as it depends on the freezing probability. For the
product aggregation rate, the gelation transition is always
suppressed, because the probability that a cluster remains reactive
decays exponentially with the number of merger events.

There are many open questions raised by this study. For example, it
will be interesting to investigate the behavior in low-dimensional
systems where the rate equation approach no longer
holds. Additionally, the exponent characterizing the mass distribution
of frozen clusters represents a novel critical exponent in percolation
processes and this should be a challenging problem below the critical
dimension.

\appendix

\section{Perturbation Analysis}
\label{pert}

\subsection{$\alpha\uparrow 1$}

To investigate the behavior in the time domain $1<\tau<\alpha^{-1}$,
we make the transformation $\tau=1+(\alpha^{-1}-1)\,x$ with $0\leq
x\leq 1$. The governing equations (\ref{mst-eq}) become 
\begin{subequations}
  \begin{align}
    \frac{d m}{dx}&= - \frac{s}{1-x}
    \label{mx-eq}\\
    \frac{d s}{dx}&= - \frac{s}{1-x} 
                - \frac{s(m-s)}{\alpha\left(1-x-\frac{s}{1-\alpha}\right)-sx}.
    \label{sx-eq}
  \end{align}
\end{subequations}
The initial conditions are now $m(0)=s(0)=1-\alpha$. When
$\alpha\uparrow 1$, we perform perturbation analysis using the small
parameter $\epsilon=1-\alpha$. Given the initial conditions
$m(0)=s(0)=\epsilon$, we write
\begin{equation}
\label{series}
m=\epsilon F_1+\epsilon^2 F_2 +\cdots\qquad
s=\epsilon G_1+\epsilon^2 G_2+\cdots. 
\end{equation}
Substituting these expansions into the governing equations
(\ref{mx-eq})--(\ref{sx-eq}) and keeping only dominant, linear in
$\epsilon$, terms we obtain
\begin{subequations}
  \begin{align}
\frac{d F_1}{dx}&= - \frac{G_1}{1-x}\,,\\
\frac{d G_1}{dx}&= - \frac{G_1}{1-x}.
  \end{align}
\end{subequations}
The initial conditions are $F_1(0)=G_1(0)=1$. Solving these equations, we find 
\begin{equation}
\label{FG-sol}
F_1=G_1=1-x.
\end{equation}
Thus, the gel mass vanishes to first order in $\epsilon$, and we
should consider the second order terms.  The second order terms are
coupled according to 
\begin{subequations}
  \begin{align}
    \frac{d F_2}{dx}&= - \frac{G_2}{1-x}
    \label{F2}\\
    \frac{d G_2}{dx}&= - \frac{G_2}{1-x}+(1-x)\,\frac{F_2-G_2}{G_2+x(1-x)}.
    \label{G2}
  \end{align}
\end{subequations}
The boundary conditions are $F_2(0)=G_2(0)=0$. We seek the non-trivial
solution with the following derivatives at the origin: $F_2'(0)=0$,
and $G_2'(0)=-2$.  

To find the probability that multiple gels are produced, we notice
that the Poisson distribution (\ref{pt}) is uniform in terms of the
transformed time, $P(x)=1$ and therefore, the probability of forming
multiple gels is simply $P_{\rm mult}=\int_0^{x_*}
dx\,P(x)=x_*$. Also, the condition (\ref{cond}) becomes
$s(x_*)=\alpha(1-\alpha)(1-x_*)$.  Substituting the perturbative
expansion $\epsilon(1-x_*)+\epsilon^2
G_2(x_*)=\epsilon(1-\epsilon)(1-x_*)$, this condition becomes
$G_2(x_*)=x_*-1$. It also implies $G_2(1)=0$.  Numerical integration
yields $P_{\rm mult}=x_*=0.45081$.

To find the maximal gel mass, we write $g=m-s=\epsilon^2 (F_2-G_2)$
and use $G_2(1)=0$. Thus, the maximal gel mass $g_{\rm max}=g(x=1)$ is
\begin{equation}
g_{\rm max}=(1-\alpha)^2 F_2(1).
\end{equation}
Integrating these equations numerically, we determine
$F_2(1)=1.303892$. Moreover, the perturbation analysis shows that in
general, the gel mass vanishes quadratically close to the critical
freezing rate, $g\propto (1-\alpha)^2$.

\subsection{$\alpha\downarrow 0$}

When freezing is slow, we may drop the freezing loss term $-\alpha
R_k$ from the master equation (\ref{Rkt}) as done in section III. The
problem therefore reduces to the no-freezing case where $u=\tau=t$ and
thus, the generating function (\ref{rzt}) is 
$\mathcal{R}(z,t)=t^{-1}G(z+\ln t-t)$. Invoking the identity
$Ge^{-G}=e^z$, the sol mass $s=\mathcal{R}(z=0)$ satisfies
$s=e^{-(1-s)t}$. In the long time limit,
\begin{equation}
\label{st-exp}
s(t)=e^{-t}+t\,e^{-2t}+\ldots
\end{equation}
{}From the one-gel criterion (\ref{cond}) and (\ref{st-exp}) we obtain
$t_*=\frac{1}{1-\alpha}\ln \frac{1}{\alpha}$.  Finally, using $P_{\rm
mult}=1-\exp[-\alpha(t_*-1)]$, we find the multiple-gel probability:
$P_{\rm mult}\simeq -\alpha\ln(e\,\alpha)$.

To leading order, the total mass $m$ remains constant. Using the exact
governing equation $dm/dt=-\alpha s$, we derive the first order
correction:
\begin{equation}
m(t)=1-\alpha-\alpha\int_1^t dt'\,s(t').
\end{equation}
The maximal mass of the gel is therefore
\begin{equation}
g_{\rm max}=1-B\alpha,\quad B=1+\int_1^{\infty}dt\, s(t).
\end{equation}
Using $t=-(1-s)^{-1}\ln s$ we change the integration variable and then
transform the integral:
\begin{eqnarray}
B=1+\int_0^1 ds\, s\left[\frac{1}{s(1-s)}+\frac{\ln s}{(1-s)^2}\right].
\end{eqnarray}
Performing the integration, we find $B=\pi^2/6$. 

\end{document}